\documentclass[a4paper,11pt]{article}
\usepackage[utf8]{inputenc}
\usepackage[]{units}
\usepackage[margin = 2.5cm]{geometry}
\usepackage{mathrsfs}
\usepackage{multirow}
\usepackage{multicol}
\usepackage{comment}
\usepackage{amsmath}
\usepackage{extarrows}
\usepackage{mathtools}
\usepackage{amsfonts}
\usepackage{amssymb}
\usepackage{enumerate}
\usepackage{diagbox}
\usepackage{amsthm}
\usepackage{bm}
\usepackage{tabularx}
\usepackage{float}
\usepackage{threeparttable}
\usepackage{graphicx}
\usepackage{stackengine}
\usepackage{enumitem}
\usepackage{cancel}
\usepackage{hyperref}

\usepackage[svgnames]{xcolor}
\usepackage[labelfont=bf]{caption}
\usepackage{subcaption}
\usepackage{amsmath}
\usepackage{blindtext}
\usepackage{wrapfig}
\usepackage{rotating}
\usepackage{caption}
\usepackage{subcaption}
\usepackage{booktabs}
\usepackage{pdflscape} 
\usepackage{xltabular}
\usepackage{titling}
\predate{}
\postdate{}
\usepackage{authblk}

\usepackage{multirow}
\usepackage{lineno, blindtext}
\usepackage{float}
\hypersetup{
	colorlinks=true,
	linktocpage=true,
	pdfstartpage=1,
	pdfstartview=FitV, 
	breaklinks=true, 
	pdfpagemode=UseNone,
	pageanchor=true, 
	pdfpagemode=UseOutlines,
	plainpages=false, 
	bookmarksnumbered,
	bookmarksopen=true,
	bookmarksopenlevel=1,
	hypertexnames=true,
	pdfhighlight=/O,
	linkcolor=black,
	citecolor=blue,
	urlcolor=blue,
	pdfsubject={},
	pdfkeywords={},
	pdfcreator={pdfLaTeX}
}

\usepackage[authoryear,round]{natbib}

\hypersetup{
    colorlinks=true,
    linkcolor=black,
    citecolor=blue,
    urlcolor=blue,
}

\title{\bf{Order of Addition in Orthogonally Blocked Mixture and Component-Amount Designs}}

\author[1]{Taha Hasan}
\author[2]{Touqeer Ahmed\thanks{Corresponding author: touqeer.ahmed@ensai.fr}} 

\affil[1]{\textit{Department of Statistics, Islamabad Model College for Boys, F-10/4, Islamabad, Pakistan}}
\affil[2]{\textit{CREST, ENSAI, University of Rennes, France}}

\date{} 

\begin{document}
\maketitle

\begin{abstract}
Mixture experiments often involve process variables, such as different chemical reactors in a laboratory or varying mixing speeds in a production line. Organizing the runs in orthogonal blocks allows the mixture model to be fitted independently of the process effects, ensuring clearer insights into the role of each mixture component. Current literature on mixture designs in orthogonal blocks ignores the order of addition of mixture components in mixture blends. This paper considers the order of addition of components in mixture and mixture-amount experiments, using the variable total amount taken into orthogonal blocks. The response depends on both the mixture proportions or the amounts of the components and the order of their addition. Mixture designs in orthogonal blocks are constructed to enable the estimation of mixture or component-amount model parameters and the order-of-addition effects. The G-efficiency criterion is used to assess how well the design supports precise and unbiased estimation of the model parameters. The fraction of the Design Space plot is used to provide a visual assessment of the prediction capabilities of a design across the entire design space.  
\end{abstract}


\section{Introduction }
\indent In product formulation, we often observe a combination of various ingredients forming a mixture. For instance, in the pharmaceutical industry, different drugs are combined in specific proportions to create an effective medication. The main focus of a mixture experiment is to determine the optimal formulation of a mixture, aiming to achieve the best possible response from a particular combination of ingredients.
An experiment where the response variable $y$ depends on the proportions $x_i$ of the components making up a mixture blend is known as a mixture experiment. The total amount of mixture is fixed. The mixture experiments have the constraints as           
 $$0 \leq x_i \leq 1 \hspace{0.2cm} \text{and} \hspace{0.2cm} 
 \sum_{i=1}^m x_i = 1 $$

The factor space of the mixture experiment in $m$ components is a $(m – 1)$ dimensional simplex $S^{m-1}$. When there are certain constraints on each mixture component, the resulting factor space is a part of the simplex $S^{m-1}$. The $n$ points in the simplex $S^{m-1}$ or in its sub-region generate a mixture design with $n$ runs.

In some situations, such as tablet formulation, the total amount also affects the response. Such experiments are called component-amount experiments, where we consider several levels of total amount. Some applications of such experiments are drug formulation, air pollution, food taste, etc. Sometimes, the mixture blends are not observed under homogeneous conditions, e.g., different shifts of plants running the experiment, different vendors supplying raw material, separate laboratories performing the experiments, etc. This lack of homogeneity is handled by making the groups of homogeneous mixture blends called blocks. These blocks are said to be orthogonal if the estimates of the blending properties in the fitted model are uncorrelated with the effects of the blocks. In our discussion, we used orthogonal blocks for both mixture and component-amount experiments.

Many physical phenomena observed in science and engineering are influenced by the order in which $m$ components or materials are added. Perhaps the first famous order-of-addition (OofA) experiment performed by \citet{fisher1971design} was tasting tea by a lady who said that she could differentiate whether tea or milk was first added to her cup. The experiment consisted of four replications of each tea order $\rightarrow$ milk and milk $\rightarrow$ tea. The OofA experiments have found wide application in nutritional science \citep{karim2000effects}, chemical experiments \citep{jiang2014sequential}, drug therapy \citep{ding2015optimized}, and in many other fields.  
 
There is limited literature on OofA mixture experiments. \citet{rios2022order} introduced OofA mixture experiments and developed designs to account for the interaction between the order of addition and mixture proportions. However, they did not address OofA mixture experiments taken into orthogonal blocks. In this contribution research, we consider the order of addition of mixture proportions or component amounts in each blend and bring it into orthogonal blocks. We assume $m!$ possible combinations of these mixture proportions/amounts are taken into orthogonal blocks of equal sizes. It is called 
OofA problem in a mixture and component-amount experiment within orthogonal blocks. The response variable $y$ depends on the order in which the component's proportions are added in a composition and the interaction effect between the order and component amounts. We generate designs for OofA mixture and component-amount experiments within orthogonal blocks based on the Latin square blocking scheme.

The paper is organized as follows: Section \ref{sec:models_for_mixture_experiment} details the models for mixture experiments 
including orthogonal blocking constraints and optimal designs for mixture experiments in orthogonal blocks. Section \ref{order_of_addition_experiment} explains OofA experiments and models for OofA experiments in orthogonal blocks. Section \ref{OofA_designs_for_mixture_experiments_in _orthogonal_blocks} explores the generation of OofA mixture designs using the optimal designs in orthogonal blocks.  Section  \ref{application_of_OofA_orthogonal_block_mixture_designs} discusses the application of generated orthogonally blocked OofA mixture designs in two real-life examples. Section \ref{mixture-amount-experiment} describes component-amount and OofA component-amount models in orthogonal blocks, D-optimal orthogonal block design for the component-amount model, and generation of OofA component-amount designs in orthogonal blocks. Section \ref{conclusion} concludes the study.
 

\section{Models for mixture experiment}
\label{sec:models_for_mixture_experiment}
In a mixture composition, the proportions of the components have a functional relationship with the response of interest. Experts aim to understand this relationship to identify the optimal mixture composition. The proportions of the components may not vary independently from one another. \citet{scheffe1958experiments, scheffe1963simplex} introduced polynomial canonical mixture models alongside of Simplex-lattice and simplex centroid designs. The quadratic form of Scheffe’s model for the expected response in $m$ components is
\begin{eqnarray}
    E(y)&=&\sum_{i=1}^m \beta_{i} x_i + \sum_{i=1}^{m-1}\sum_{j=i+1}^m \beta_{ij}x_ix_j.
\end{eqnarray}
\citet{draper1998mixture} introduced Kronecker polynomial models (K-models) based on the Kronecker algebra of vectors and matrices. The quadratic K-model is given as
\begin{eqnarray}
    E(y)&=&\sum_{i=1}^m \theta_{ii} x_i^{2} + \sum_{i=1}^{m-1}\sum_{j=i+1}^m \theta_{ij}x_ix_j.
\end{eqnarray}
Many other mixture models, such as the Cox, Slack-variable, Log-contrast, Octane-blending, and homogeneous blending models, are available in the literature. 
\subsection{Orthogonal Blocking in Mixture Experiments}
\label{subsec:orthogonal_blocking_in_mixture_experiments}
In many situations, conducting all experimental runs under similar conditions is impractical, making it necessary to split the experiment into blocks. For easier interpretation, it is preferable if the model terms can be estimated independently of the block effects, which is achieved through orthogonal blocking.
\citet{murty1966} developed orthogonal block designs for mixture experiments using a transformation technique. \citet{nigam1970block,nigam1976correction} defined combinatorial tools and drew methods for constructing orthogonal block designs for second-order canonical polynomial models but the conditions were quite restrictive. \citet{john1984experiments} provided necessary and sufficient conditions for mixtures within orthogonal blocks. \citet{john1984experiments} used $m \times m$ Latin squares for the construction of orthogonal block designs in $m$ mixture components using \citet{scheffe1958experiments,scheffe1963simplex} quadratic mixture model. For $u$th blend the quadratic Scheffe’s mixture model in terms of blocking variable $v_u$, having $p$ blends in each block with $n$ =$2p$ is
\begin{eqnarray}
    Y_u &=& \sum_{i=1}^m \beta_i x_{ui} + \sum_{i=1}^{m-1} \sum_{j=i+1}^m \beta_{ij} x_{ui} x_{uj} +\gamma v_u + e_u, \quad u = 1, 2, \dots, n, 
\end{eqnarray}
where $v_u$  = -1 for the blends in the first block and $v_u$  = +1 for the blends in second block. In matrix form, the model can be written as
$$E(\mathbf{y}) = \mathbf{X}\beta + \boldsymbol{\gamma}\mathbf{v},$$ 
where $\mathbf{X}$ is the matrix belongs to the mixture part, $\mathbf{y}$ is a column vector of observed values and $\mathbf{v}$ is a column vector of blocking variable. The blocks are orthogonal if  
$$\mathbf{X}^{T} \mathbf{v} = 0$$
 and the following necessary and sufficient conditions for orthogonal blocking are satisfied 
$$
\sum_{u=1}^{n_w} x_{ui} = k_i, \quad \sum_{u=1}^{n_w} x_{ui}x_{uj} = k_{ij}, \quad w = 1,2, \quad i < j, \quad i, j = 1, 2, \dots, m.
$$
A full mixture of centroids is added in each block to address the singularity problem in the model design matrix. When we use the Quadratic K-model, the necessary and sufficient conditions for orthogonal blocking are 
$$
\sum_{u=1}^{n_w} x_{ui}^2 = k_{ii}, \quad \sum_{u=1}^{n_w} x_{ui}x_{uj} = k_{ij}, \quad w = 1,2, \quad i < j, \quad i, j = 1, 2, \dots, m, 
$$
where $k_i$ , $k_{ii}$ and  $k_{ij}$ are constants.

\subsection{Optimal Mixture Designs in Orthogonal Blocks}
\label{optimal_mixture_designs-in-orthogonal-blocks}
Optimal designs are experimental designs that maximize efficiency based on specific statistical criteria. Model estimation allows parameters to be estimated with minimum variance and without bias. The optimality of a design is determined by the statistical model and evaluated using criteria related to the estimator's variance-covariance matrix. Common criteria include D-, A-, E-, and G-optimality. This discussion focuses on D- and A-optimal designs in two orthogonal blocks using the quadratic Scheffé mixture model and the quadratic K-mixture model.
 The D-optimality criterion suggests that one should reduce uncertainty in model coefficient estimates and seek to minimize $ \left| (X^{T}X)^{-1} \right|$ or equivalently maximize the determinant of the information matrix of the design. It is associated with minimizing the volume of the confidence ellipsoid for the estimated coefficients. 
 
 The A-optimality criterion is used to search the minimum trace of the inverse of the information matrix $ \left| (X^{T}X)\right|$. This ensures the minimum average variance of the parameter estimates. A-optimal designs generally efficiently estimate the model parameters, although they may not be the best designs for D-optimality. \citet{czitrom1988mixture}  discussed D-optimal Latin square-based orthogonally blocked designs in two blocks for three components for Scheffe’s quadratic mixture model.  The D-optimal design in two orthogonal blocks with a single pair of Latin squares in each block is given in Table~\ref{Tab:1}.

\begin{table}[htbp]
\centering
\caption{D-optimal design in two orthogonal blocks for Scheffe’s quadratic model.}
\label{Tab:1}
\begin{tabular}{cccc|cccc}
\toprule
\multicolumn{4}{c|}{Block I} & \multicolumn{4}{c}{Block II} \\
\midrule
Run & $x_1$ & $x_2$ & $x_3$ & Run & $x_1$ & $x_2$ & $x_3$ \\
\midrule
1  & 0.168 & 0.832 & 0     & 5 & 0.168 & 0     & 0.832 \\
2  & 0.832 & 0     & 0.168 & 6 & 0.832 & 0.168 & 0     \\
3  & 0     & 0.168 & 0.832 & 7 & 0     & 0.832 & 0.168 \\
4  & 0.333 & 0.333 & 0.334 & 8 & 0.333 & 0.333 & 0.334 \\
\bottomrule
\end{tabular}
\end{table}

\citet{aggarwal2004orthogonal} derived D-, A- and E-optimal Latin square based orthogonally blocked designs in two blocks with three and four components, for a quadratic K-model. We only consider A-optimal design in two orthogonal blocks with three components and a single pair of Latin square in each block. The design is given blelow in Table \ref{Tab:2}.
\begin{table}[htbp]
\centering
\caption{A-optimal design in two orthogonal blocks for quadratic K-model.}
\label{Tab:2}
\begin{tabular}{cccc|cccc}
\toprule
\multicolumn{4}{c|}{Block I} & \multicolumn{4}{c}{Block II} \\
\midrule
Run & $x_1$ & $x_2$ & $x_3$ & Run & $x_1$ & $x_2$ & $x_3$ \\
\midrule
1  & 0.239 & 0.761 & 0     & 5 & 0.239 & 0     & 0.761 \\
2  & 0.761 & 0     & 0.239 & 6 & 0.761 & 0.239 & 0 \\
3  & 0     & 0.239 & 0.761 & 7 & 0     & 0.761 & 0.239 \\
4  & 0.333 & 0.333 & 0.334 & 8 & 0.333 & 0.333 & 0.334 \\
\bottomrule
\end{tabular}
\end{table}
\section{Order-of-Addition experiments}
\label{order_of_addition_experiment}
The current literature on OofA experiments focuses primarily on the Pair-Wise Ordering (PWO) model, first introduced by \citet{van1995design}. The model was formally named the PWO model by \citet{voelkel2019design}. For a set of \( m \) components, their permutation is denoted by $\mathbf{c} = (c_1, c_2, \dots, c_m)^T $.
Let \( \mathcal{P} \) be the set of all pairs \( (j, k) \) where \( 1 \leq j < k \leq m \).The PWO factor \( z_{jk}(\mathbf{c}) \) is defined as:  
\begin{equation*}
    z_{jk}(\mathbf{c})=
    \begin{cases}
          1 &\text{if $j$ precedes $k$ in $c$}\\
         -1 & \text{if $j$ precedes $j$ in $c$}
    \end{cases},
\end{equation*}
hence if $\mathbf{c}=(2, 1, 3)$ then \( z_{12}(\mathbf{c})=-1 \), \( z_{13}(\mathbf{c})= 1 \) and  \( z_{23}(\mathbf{c})=1 \).

The PWO model for the expected response $E(c)$ is given as
$$E(c) =\beta_0 +\sum_{jk\in S}z_{jk}(\mathbf{c}) \beta_{jk}$$

The parameter $\beta_{jk}$ shows how the pair-wise order of components $j$ and $k$ affect the response. For finding the optimal order \citet{lin2019order} discussed the topological sorting methods to be used in the PWO model. There are several research papers on the Optimality of PWO designs. See for instance, \citet{peng2019design, winker2020construction, Zhao2021}.
\subsection{Models for OofA Mixture Experiment in orthogonal blocks}
\label{models_for_OofA_mixture_experiment_in_orthogonal_blocks}
Three models for mixture experiments were discussed in section \ref{sec:models_for_mixture_experiment} without considering the order-of-addition of components in the blends. Now we revise the models for OofA mixture experiments in orthogonal blocks. We use the modified PWO notations given by \citet{rios2022order}. Define $(m-1)$ dimensional simplex $S$ and $\mathcal{P}$ be the set of all pairs $(j, k)$ where $1 \leq j <k \leq m$. To get the full design matrix for fitting the mixtur models we define PWO factor $z_{jk} (x, \mathbf{c})$
\begin{equation*}
	 z_{jk}(x,\mathbf{c}) =
	\begin{cases} 
		1 &x_j, x_i \neq 0 \hspace{0.1cm}  \text{$j$ is before $k$ in $c$}\\
  0 & x_j = 0  \hspace{0.1cm} or \hspace{0.1cm}x_i=0\\
  - 1 & x_j, x_i \neq 0  \hspace{0.1cm} \text{$k$ is before $j$ in $c$.}\\
	\end{cases}
\end{equation*} 
If we assume that the order-of-addition of components affects the mixture response, the general model for pair-wise ordering and mixture component effects includes mixture-order interaction term is
$$y = f(x,z)+\epsilon,$$
where the error term $\epsilon \sim N(0, \sigma^2)$. The quadratic form of Scheffe's model for the expected response, including $m$ components with mixture proportions and PWO variables effect along with the mixture-order interactions is given as
\begin{eqnarray}\label{scheffe}
    E(y)&=&\sum_{i=1}^m \beta_{i} x_i + \sum_{i=1}^{m-1}\sum_{j=i+1}^m \beta_{ij}x_ix_j + \sum_{k<l}\delta_{k}z_{kl}+\sum_{i=1}^m\sum_{k<l,i=k,l}\lambda_{kl}^ix_iz_{kl}.
\end{eqnarray}
Let $m$ mixture blends are arranged into two blocks. For the $u$th blend the quadratic Scheffe’s mixture model with PWO variables and in terms of blocking variable $v_u$ is

\begin{eqnarray} \label{model:5} 
    Y_u &=& \sum_{i=1}^m \beta_i x_{ui} + \sum_{i=1}^{m-1} \sum_{j=i+1}^m \beta_{ij} x_{ui} x_{uj} \nonumber \\
    && + \sum_{k<l} \delta_k z_{u(kl)} + \sum_{i=1}^m \sum_{k<l,i=k,l} \lambda_{kl}^i x_{ui} z_{u(kl)} + \gamma v_u + e_u
\end{eqnarray}
\noindent with $u = 1, 2, \dots, n.$
As mentioned earlier in \ref{subsec:orthogonal_blocking_in_mixture_experiments} about the necessary and sufficient conditions for orthogonal blocking with respect to different types of mixture models. Those conditions ensure the independence of coefficient estimates of mixture effects from block effects in the model. The model now includes the order-of-addition effect. Ensuring block effects are independent of the coefficient estimates for PWO variable effects, we define two more necessary and sufficient conditions for orthogonal blocking.

$$
\sum_{u=1}^{n_w} z_{u(kl)} = f_{kl}, \quad \sum_{u=1}^{n_w} x_{ui}z_{u(kl)} = f_{i(kl)}, \quad w = 1,2, \quad i = 1, 2, \dots, m
$$
where $f_{kl}$ and $f_{i(kl)}$ are constants.
The quadratic K-model including PWO variables effect and its interaction effect with mixture proportions is
\begin{eqnarray}
    E(y)&=&\sum_{i=1}^m \theta_{ii} x_{i}^2 + \sum_{i=1}^{m-1}\sum_{j=i+1}^m \theta_{ij}x_ix_j + \sum_{k<l}\delta_{k}z_{kl}+\sum_{i=1}^m\sum_{k<l,i=k,l}\lambda_{kl}^ix_iz_{kl}
\end{eqnarray}
The response model for $u$th blend in any orthogonal block $w$ is
\begin{eqnarray}  
    Y_u &=& \sum_{i=1}^m \theta_{ii} x_{ui}^2 + \sum_{i=1}^{m-1} \sum_{j=i+1}^m \theta_{ij} x_{ui} x_{uj} \nonumber \\
    && + \sum_{k<l} \delta_k z_{u(kl)} + \sum_{i=1}^m \sum_{k<l,i=k,l} \lambda_{kl}^i x_{ui} z_{u(kl)} + \gamma v_u + e_u
\end{eqnarray}
\noindent with $u = 1, 2, \dots, n.$
\section{OofA designs for mixture experiments in orthogonal blocks}
\label{OofA_designs_for_mixture_experiments_in _orthogonal_blocks}
Now, we construct OofA mixture designs in two orthogonal blocks using the optimal orthogonally blocked mixture 
designs given in section \ref{optimal_mixture_designs-in-orthogonal-blocks}, for the different mixture models given in section \ref{models_for_OofA_mixture_experiment_in_orthogonal_blocks}.
\subsection{Construction of OofA mixture design in two orthogonal blocks using Scheffe’s D-optimal orthogonally 
         blocked design}
In section \ref{optimal_mixture_designs-in-orthogonal-blocks}, we discussed a three-component D-optimal design in two orthogonal blocks for Scheffe’s quadratic model, derived by \citet{czitrom1988mixture}. Using this design, we generate a three-component OofA mixture design in two orthogonal blocks containing mixture proportions ($x_1$, $x_2$, $x_3$) and the PWO variable ($z_{12}$, $z_{13}$ ,$z_{23}$), where $z_{jk}$ is defined in section \ref{models_for_OofA_mixture_experiment_in_orthogonal_blocks}. For fitting OofA Scheffe’s quadratic mixture model (\ref{model:5}) in two orthogonal blocks, the design matrix is given in Table~\ref{Tab:3}.
\begin{table}[htbp]
\centering
\caption{D-optimal design in two orthogonal blocks with interaction terms for quadratic model.}
\label{Tab:3}
\begin{tabular}{ccccccc|ccccccc}
\toprule
\multicolumn{7}{c|}{Block I} & \multicolumn{7}{c}{Block II} \\
\midrule
Run & $x_1$ & $x_2$ & $x_3$ & $z_{12}$ & $z_{13}$ & $z_{23}$ & Run & $x_1$ & $x_2$ & $x_3$ & $z_{12}$ & $z_{13}$ & $z_{23}$ \\
\midrule
1  & 0.168 & 0.832 & 0     & 1  & 0  & 0  & 13 & 0.168 & 0    & 0.832 & 0  & 1  & 0  \\
2  & 0.168 & 0.832 & 0     & -1 & 0  & 0  & 14 & 0.168 & 0    & 0.832 & 0  & -1 & 0  \\
3  & 0.832 & 0     & 0.168 & 0  & -1 & 0  & 15 & 0.832 & 0.168 & 0    & -1 & 0  & 0  \\
4  & 0.832 & 0     & 0.168 & 0  & 1  & 0  & 16 & 0.832 & 0.168 & 0    & 1  & 0  & 0  \\
5  & 0     & 0.168 & 0.832 & 0  & 0  & 1  & 17 & 0     & 0.832 & 0.168 & 0  & 0  & -1 \\
6  & 0     & 0.168 & 0.832 & 0  & 0  & -1 & 18 & 0     & 0.832 & 0.168 & 0  & 0  & 1  \\
7  & 0.333 & 0.333 & 0.334 & 1  & 1  & 1  & 19 & 0.333 & 0.333 & 0.334 & 1  & 1  & 1  \\
8  & 0.333 & 0.333 & 0.334 & 1  & 1  & -1 & 20 & 0.333 & 0.333 & 0.334 & 1  & 1  & -1 \\
9  & 0.333 & 0.333 & 0.334 & 1  & -1 & -1 & 21 & 0.333 & 0.333 & 0.334 & 1  & -1 & -1 \\
10 & 0.333 & 0.333 & 0.334 & -1 & 1  & 1  & 22 & 0.333 & 0.333 & 0.334 & -1 & 1  & 1  \\
11 & 0.333 & 0.333 & 0.334 & -1 & -1 & 1  & 23 & 0.333 & 0.333 & 0.334 & -1 & -1 & 1  \\
12 & 0.333 & 0.333 & 0.334 & -1 & -1 & -1 & 24 & 0.333 & 0.333 & 0.334 & -1 & -1 & -1 \\
\bottomrule
\end{tabular}
\end{table}

\subsection{Construction of OofA orthogonal block mixture design using A-optimal design for K-model in two orthogonal blocks}
Orthogonal block A-optimal design in three components, for quadratic K-model, is discussed in section \ref{optimal_mixture_designs-in-orthogonal-blocks} (see Table~\ref{Tab:2}). We use it to generate an orthogonal block OofA mixture design in two blocks for the quadratic K-model. The design contains mixture proportions($x_1$, $x_2$, $x_3$) and the PWO variable ($z_{12}$, $z_{13}$ ,$z_{23}$) with mixture-order interaction effect. The design matrix for a quadratic K-model with PWO effect is given in Table~\ref{Tab:4}.
\begin{table}[htbp]
\centering
\caption{A-optimal design in two orthogonal blocks with interaction terms for quadratic model.}
\label{Tab:4}
\begin{tabular}{ccccccc|ccccccc}
\toprule
\multicolumn{7}{c|}{Block I} & \multicolumn{7}{c}{Block II} \\
\midrule
Run & $x_1$ & $x_2$ & $x_3$ & $z_{12}$ & $z_{13}$ & $z_{23}$ & Run & $x_1$ & $x_2$ & $x_3$ & $z_{12}$ & $z_{13}$ & $z_{23}$ \\
\midrule
1  & 0.239 & 0.761 & 0     & 1  & 0  & 0  & 13 & 0.239 & 0    & 0.761 & 0  & 1  & 0  \\
2  & 0.239 & 0.761 & 0     & -1 & 0  & 0  & 14 & 0.239 & 0    & 0.761 & 0  & -1 & 0  \\
3  & 0.761 & 0     & 0.239 & 0  & -1 & 0  & 15 & 0.761 & 0.239 & 0    & -1 & 0  & 0  \\
4  & 0.761 & 0     & 0.239 & 0  & 1  & 0  & 16 & 0.761 & 0.239 & 0    & 1  & 0  & 0  \\
5  & 0     & 0.239 & 0.761 & 0  & 0  & 1  & 17 & 0     & 0.761 & 0.239 & 0  & 0  & -1 \\
6  & 0     & 0.239 & 0.761 & 0  & 0  & -1 & 18 & 0     & 0.761 & 0.239 & 0  & 0  & 1  \\
7  & 0.333 & 0.333 & 0.334 & 1  & 1  & 1  & 19 & 0.333 & 0.333 & 0.334 & 1  & 1  & 1  \\
8  & 0.333 & 0.333 & 0.334 & 1  & 1  & -1 & 20 & 0.333 & 0.333 & 0.334 & 1  & 1  & -1 \\
9  & 0.333 & 0.333 & 0.334 & 1  & -1 & -1 & 21 & 0.333 & 0.333 & 0.334 & 1  & -1 & -1 \\
10 & 0.333 & 0.333 & 0.334 & -1 & 1  & 1  & 22 & 0.333 & 0.333 & 0.334 & -1 & 1  & 1  \\
11 & 0.333 & 0.333 & 0.334 & -1 & -1 & 1  & 23 & 0.333 & 0.333 & 0.334 & -1 & -1 & 1  \\
12 & 0.333 & 0.333 & 0.334 & -1 & -1 & -1 & 24 & 0.333 & 0.333 & 0.334 & -1 & -1 & -1 \\
\bottomrule
\end{tabular}
\end{table}

\section{Application of OofA orthogonal block mixture designs}
\label{application_of_OofA_orthogonal_block_mixture_designs}
Building on the OofA mixture designs generated in the previous sections where we utilized D-optimal designs for Scheffé's quadratic model and A-optimal designs for the quadratic K-model within two orthogonal blocks, incorporating the order-of-addition effect, we now apply these orthogonal block OofA mixture designs to real-life examples of mixture experiments. This application demonstrates the practical utility and effectiveness of the designs in capturing the complex interactions between component proportions and their order of addition.
\subsection{Example 1: Yarn forming in a Textile factory}
\label{Example_1}
We take an example from \citet{cornell2002experiments} of yarn forming using three constituents: Polyethylene ($x_1$), Polystyrene ($x_2$), and Polypropylene ($x_3$). The three ingredients were blended together, and the resulting material was spun into yarn. We are interested in the strength of the yarn (the response variable) and it is formed on two different types of machines. The machine type is considered as a blocking or process variable. So, we have a single process variable $v$ with two levels 
($v$ = -1, $v$ =+1). The blends are distributed into two blocks which are orthogonal to each other. The objective of the analysis includes how the order of addition and proportions in this orthogonal block mixture experiment impacted the response.  We applied the OofA Scheffe's mixture model \eqref{scheffe} to the design matrix given in Table~\ref{Tab:3}. For model identifiability, we omit the mixture-order interactions $x_1z_{13}$, $x_2z_{12}$, $x_3z_{23}$ and only consider $x_1z_{12}$, $x_2z_{23}$, $x_3z_{31}$. 

The design analysis shows a maximum prediction variance of 0.922, indicating fairly consistent prediction accuracy, while an average prediction variance of 0.542 suggests moderate uncertainty. A G-efficiency of 58.8\% reflects reasonable efficiency. Overall, the design is well-constructed, balancing the number of runs with precise parameter estimates.

\begin{table}[htbp]
\centering
\caption{Power at 5\% alpha level for effects.}
\label{Tab:5}
\begin{tabular}{cccccc}
\toprule
Terms & Std. Error & 2 Std. Dev & Terms & Std. Error & 2 Std. Dev \\
\midrule
$x_1$      & 0.81 & 50.4\% & $x_1x_2$     & 1.10 & 13.2\% \\
$x_2$      & 0.81 & 50.4\% & $x_1x_3$     & 1.10 & 13.2\% \\
$x_3$      & 0.81 & 50.4\% & $x_2x_3$     & 1.10 & 13.2\% \\
$z_{12}$   & 0.32 & 81.3\% & $x_1z_{12}$  & 0.69 & 26.0\% \\
$z_{13}$   & 0.32 & 81.3\% & $x_1z_{13}$  & 0.69 & 26.0\% \\
$z_{23}$   & 0.32 & 81.3\% & $x_2z_{23}$  & 0.69 & 26.0\% \\
\bottomrule
\end{tabular}
\end{table}
Power analysis at the 5\% level for effects is given in Table~s\ref{Tab:5}. The three mixture ingredients $x_1$, $x_2$, $x_3$ and the order-of-addition factors $z_{12}$, $z_{13}$, $z_{23}$ are all significant. Order-of-addition factors show higher significance than others. This suggests that order-of-addition variables have a more pronounced effect on the response variable. All the two-factor interactions are significant. The interactions $x_1z_{12}$, $x_2z_{23}$, $x_3z_{13}$ show moderate significance, with power at 2 times standard deviations reaching up to 26.0\%. This indicates that mixture-order interactions have a noticeable effect. 
\subsection{Example 2: Vinyl production for automobile seat covers}\label{Example 2}
 For producing vinyl used in automobile seat covers, mixture formulations involve several types of components (plasticizers, stabilizers, lubricants, drying agents, and resins). A preliminary experiment was set up to study three plasticizers: Di(2-ethylhexyl)-phthalate, Di(2-ethylhexyl)- adipate, and Epoxidized soybean oil, where the effect of other components was fixed. The measured response $y$ is the vinyl thickness value. The experiment is run under one blocking variable, i.e., the temperature of drying at two levels (Temp$_{\text{low}} = 25^\circ\text{C}$, Temp$_{\text{high}} = 70^\circ\text{C}$). Different temperatures are used to assess the impact of temperature on vinyl thickness.
The design analysis reveals a maximum prediction variance of 0.941, indicating fairly consistent prediction accuracy. An average prediction variance of 0.542 reflects overall prediction stability, and a G-efficiency value of 58.8\% reflects reasonable design efficiency. 

\begin{table}[htbp]
\centering
\caption{Power at 5\% alpha level for effects.}
\label{Tab:6}
\begin{tabular}{cccccc}
\toprule
Terms & Std. Error & 2 Std. Dev & Terms & Std. Error & 2 Std. Dev \\
\midrule
$x_1^2$    & 0.62 & 83.4\% & $x_1x_2$     & 0.88 & 18.1\% \\
$x_2^2$    & 0.62 & 83.4\% & $x_1x_3$     & 0.88 & 18.1\% \\
$x_3^2$    & 0.62 & 83.4\% & $x_2x_3$     & 0.88 & 18.1\% \\
$z_{12}$   & 0.34 & 76.6\% & $x_1z_{12}$  & 0.85 & 19.1\% \\
$z_{13}$   & 0.34 & 76.6\% & $x_1z_{13}$  & 0.85 & 19.1\% \\
$z_{23}$   & 0.34 & 76.6\% & $x_2z_{23}$  & 0.85 & 19.1\% \\
\bottomrule
\end{tabular}
\end{table}
The power analysis in Table~\ref{Tab:6} at 5\% level reveals that the mixtures quadratic effects $x_1^2$, $x_2^2$, $x_3^2$ and order-of-addition effects $z_{12}$, $z_{13}$, $z_{23}$ have high probability for detecting effects. This indicates that they are all significant effects. The mixture effects and the order-of-addition effect of three plasticizers in the mixture blend significantly affect the response, i.e., vinyl thickness. The interaction of quadratic mixture effects and order-of-addition effects show moderate significance, though it is less pronounced than their individual effects.
\section{Mixture-Amount Experiments}\label{mixture-amount-experiment}
A mixture-amount experiment is performed at two or more levels of the total amount. The response is assumed to be dependent on the individual proportions of components in the blend and also on the amount of blend. A common example is the formulation of a pain relief tablet composed of different proportions of active pharmaceutical ingredients, a binder, and a disintegrant. The experimenter wishes to know the tablet weight and find the best relative proportions of the ingredients. Other areas of application are agriculture, nutrition science, chemical industry bio-engineering, and so on. The purpose of the mixture-amount experiment is to learn about the effects of the varying total amount of mixture and varying proportions of mixture components on response. 
\subsection{Models for Mixture-Amount Experiment}
The design for fitting the mixture-amount model is called the mixture-amount design, developed by \citet{piepel1985models}. The linear and quadratic expected response mixture-amount models for $m$ components is
\begin{eqnarray}
    E(y)&=&\sum_{i=1}^m \gamma_{i}^0 x_i + \sum_{i=1}^m \gamma_{i}^1 x_i A \nonumber \\
   E(y)&=&  \sum_{i=1}^m \gamma_{i}^0 x_i+ \sum_{i<j}^m \gamma_{ij}^0 x_i x_j + \sum_{l=1}^2\left(\sum_{i=1}^m \gamma_{i}^l x_i+ \sum_{i<j}^m \gamma_{ij}^l x_i x_j\right)A^l
\end{eqnarray}
This model consists of three second-order Scheff\`e polynomial forms in $m$ components, each multiplied by the powers of the total amount $A$ ($A^{0}=1$, $A$ and $A^{2}$). When $A = 0$, this model predicts a zero response. \citet{piepel1988note} modified this model to accommodate zero-amount condition. The alternative model uses the actual amounts of the components denoted by $a_1, a_2, \dots, a_m$ such that $a_1+a_2 \dots + a_m = A$. The proportions $x_i$ are related to the amounts $a_i$ through $x_i = a_i /A$ such that $a_i = x_i A$. This is called a component-amount model. The linear and quadratic expected response component-amount models are,
\begin{eqnarray}
    E(y)&=& \alpha_0 +\sum_{i=1}^m \alpha_{i} a_i \nonumber \\
    E(y)&=& \alpha_0 +\sum_{i=1}^m \alpha_{i} a_i + \sum_{i=1}^m \left(\alpha_{ii} a_i^2 + \sum_{i<j}^m \alpha_{ij} a_i a_j\right)
\end{eqnarray}
 For further discussion on mixture-amount and component-amount models, see, for instance,  \citet[pp. 403-418]{cornell2002experiments}
\subsection{Component-amount models for experiments in orthogonal blocks with OofA effect}
\label{component_amount_models_for_experiments_in_orthogonal _blocks_with_OofA_effect}
In this section, we derive component-amount models for experiments in orthogonal blocks and incorporate the order-of-addition effect. The expected response additive model for the component amount and PWO factor without interaction is    
\begin{eqnarray}
    E(y)&=&\alpha_0 +\sum_{i=1}^m \alpha_i a_{i} + \sum_{k<l} \beta_{kl} Z_{kl} 
\end{eqnarray}
For $u$th blend the linear component-model in terms of blocking variable $v_u$ with $p$ blends in each block, and $n$ =$2p$ is
\begin{eqnarray}
    y_u &=&\alpha_0 +\sum_{i=1}^m \alpha_i a_{ui} + \sum_{k<l} \beta_{kl} Z_{u(kl)} + \gamma v_u + e_u
\end{eqnarray}
When we include component-amount and order interaction, the model becomes
\begin{eqnarray}\label{order_inter}
    E(y)&=&\alpha_0 +\sum_{i=1}^m \alpha_i a_{i} + \sum_{k<l} \beta_{kl} Z_{kl} +\sum_{i=1}^m \alpha_{ii} a_{i}^2 + \sum_{i<j} \gamma_{ij} a_i a_j \nonumber \\
    &+& \sum_{i} \sum_{k<l \hspace{0.1cm}\text{or} \hspace{0.1cm}k>l, k\neq l } \delta_{i(kl)} a_i Z_{kl}
\end{eqnarray}
For the $u$th blend, the quadratic OofA component-amount model with orthogonal block effect $v$ is
\begin{eqnarray}\label{order_inter}
    y_u &=&\alpha_0 +\sum_{i=1}^m \alpha_i a_{ui} + \sum_{k<l} \beta_{kl} Z_{u(kl)} +\sum_{i=1}^m \alpha_{ii} a_{ui}^2 + \sum_{i<j} \gamma_{ij} a_{ui} a_{uj} \nonumber \\
    &+& \sum_{i} \sum_{k<l \hspace{0.1cm}\text{or} \hspace{0.1cm}k>l, k\neq l } \delta_{i(kl)} a_{ui} Z_{u(kl)}+\gamma v_u + e_u
\end{eqnarray}
\subsection{OofA Component-Amount design in two Orthogonal Blocks}
\citet{prescott2004mixture} derived orthogonal block designs for the component-amount model from the projection of orthogonally blocked mixture designs, based on Latin squares, into lower dimensions.  A three-component D-optimal design in two orthogonal blocks obtained via projection of a four components orthogonal block mixture design, given by \citet[p.224]{prescott2004mixture}, is used to generate an OofA component-amount mixture design in two orthogonal blocks. The design is given in Table \ref{Tab:7}. The design has four levels of total amount A, i.e., $A$ = 0.24, 0.75, 0.76, and 1.00, with 36 mixture blends. 
\begin{table}[t]
\centering
\caption{The OofA simplex-lattice design with \( m = 3 \) , \( l = 3 \) and with three levels of total amount \( A \).}
\label{Tab:7}
\scriptsize
\centering
\begin{tabular}{p{0.5cm} p{0.5cm} p{0.5cm} p{0.5cm} p{0.5cm} p{0.5cm} p{0.5cm} p{0.5cm} | p{0.5cm} p{0.5cm} p{0.5cm} p{0.5cm} p{0.5cm} p{0.5cm} p{0.5cm} p{0.5cm}}
  \hline
 Runs & $a_1$ & $a_2$ & $a_3$ & $z_{12}$ & $z_{13}$ & $z_{23}$ & $A$ & Runs & $a_1$ & $a_2$ & $a_3$ & $z_{12}$ & $z_{13}$ & $z_{23}$ & $A$ \\
  \hline
1  & 0    & 0    & 0.24 & 0    & 0    & 0    & 0.24 & 19 & 0    & 0.24 & 0    & 0    & 0    & 0    & 0.24 \\
2  & 0    & 0.76 & 0    & 0    & 0    & 0    & 0.76 & 20 & 0    & 0    & 0.76 & 0    & 0    & 0    & 0.76 \\
3  & 0.24 & 0    & 0.76 & 0    & 1    & 0    & 1.00 & 21 & 0.24 & 0.76 & 0    & 1    & 0    & 0    & 1.00 \\
4  & 0.24 & 0    & 0.76 & 0    & -1   & 0    & 1.00 & 22 & 0.24 & 0.76 & 0    & -1   & 0    & 0    & 1.00 \\
5  & 0.76 & 0.24 & 0    & -1   & 0    & 0    & 1.00 & 23 & 0.76 & 0    & 0.24 & 0    & -1   & 0    & 1.00 \\
6  & 0.76 & 0.24 & 0    & 1    & 0    & 0    & 1.00 & 24 & 0.76 & 0    & 0.24 & 0    & 1    & 0    & 1.00 \\
7  & 0    & 0    & 0.24 & 0    & 0    & 0    & 0.24 & 25 & 0    & 0.76 & 0.24 & 0    & 0    & -1   & 1.00 \\
8  & 0    & 0.24 & 0.76 & 0    & 0    & 1    & 1.00 & 26 & 0    & 0.76 & 0.24 & 0    & 0    & 1    & 1.00 \\
9  & 0    & 0.24 & 0.76 & 0    & 0    & -1   & 1.00 & 27 & 0    & 0    & 0.76 &      & 0    & 0    & 0.76 \\
10 & 0.24 & 0.76 & 0    & 1    & 0    & 0    & 1.00 & 28 & 0.24 & 0    & 0    & 0    &  0   & 0    & 0.24 \\
11 & 0.24 & 0.76 & 0    & -1   & 0    & 0    & 1.00 & 29 & 0.76 & 0.24 & 0    & -1   & 0    & 0    & 1.00 \\
12 & 0.76 & 0    & 0    & 0    & 0    & 0    & 0.76 & 30 & 0.76 & 0.24 & 0    & 1    & 0    & 0    & 1.00 \\
13 & 0.25 & 0.25 & 0.25 & 1    & 1    & 1    & 0.75 & 31 & 0.25 & 0.25 & 0.25 & 1    & 1    & 1    & 0.75 \\
14 & 0.25 & 0.25 & 0.25 & 1    & 1    & -1   & 0.75 & 32 & 0.25 & 0.25 & 0.25 & 1    & 1    & -1   & 0.75 \\
15 & 0.25 & 0.25 & 0.25 & 1    & -1   & -1   & 0.75 & 33 & 0.25 & 0.25 & 0.25 & 1    & -1   & -1   & 0.75 \\
16 & 0.25 & 0.25 & 0.25 & -1   & 1    & 1    & 0.75 & 34 & 0.25 & 0.25 & 0.25 & -1   & 1    & 1    & 0.75 \\
17 & 0.25 & 0.25 & 0.25 & -1   & -1   & 1    & 0.75 & 35 & 0.25 & 0.25 & 0.25 & -1   & -1    & 1   & 0.75 \\
18 & 0.25 & 0.25 & 0.25 & -1   & -1   & -1   & 0.75 & 36 & 0.25 & 0.25 & 0.25 & -1   & -1    & -1  & 0.75 \\
\hline
\end{tabular}
\end{table}
\subsection{Example 3: A Migraine relief tablet formulation}\label{Example 3}
In the formulation of migraine relief tablets, studies have shown that Sumatriptan-based formulations are widely used in treating migraine, as discussed in \citet{munija2018formulation}. The three key components include the active ingredient, Sumatriptan (API), a filler (Lactose Monohydrate), and a disintegrant (Microcrystalline Cellulose). A tablet formulation has several blocking variables that can be considered to control variability in the manufacturing process. Blocking variables helps to reduce the impact of factors that are not of primary interest in the design but may influence the outcome. Different blocking variables that may generally be considered during a tablet formulation are manufacturing equipment, batch size, operator, raw material supplier, storage conditions, environmental conditions, granulation method, compression force, granule moisture content, and coating process. For simplicity, we will focus on a single blocking variable: the granulation method at two levels, i.e., wet granulation and dry granulation. The granulation method can influence the particle size, distribution, and compressibility of the granules. Blocking ensures that these differences do not obscure the response being studied. The measured response is the crushing strength of the tablets.

In practice, the total weight of the three components in Migraine relief tablet formulation is typically up to 100 mg, i.e., A$_{\text{max}} = 100$ mg. We assume that several dosage strengths are to be formulated. We perform the experiment to observe whether the order of addition of three components and their different amounts affect the crushing strength of the tablet. We use a component-amount model incorporating PWO variables effect and block effect. The model is fitted in OofA component-amount design in two orthogonal blocks, given in Table \ref{Tab:8}. With the total amount of 100 mg, the design has four levels of $A$, i.e., 24, 75, 76, and 100, with replications 4, 4, 12, and 16, respectively. 
\begin{table}[t]
\centering
\caption{Design for OofA component-amount model in orthogonal blocks with Amax = 100 mg }
\label{Tab:8}
\scriptsize
\centering
\begin{tabular}{p{0.5cm} p{0.5cm} p{0.5cm} p{0.5cm} p{0.5cm} p{0.5cm} p{0.5cm} p{0.5cm} | p{0.5cm} p{0.5cm} p{0.5cm} p{0.5cm} p{0.5cm} p{0.5cm} p{0.5cm} p{0.5cm}}
  \hline
 Runs & $a_1$ & $a_2$ & $a_3$ & $z_{12}$ & $z_{13}$ & $z_{23}$ & $A$ & Runs & $a_1$ & $a_2$ & $a_3$ & $z_{12}$ & $z_{13}$ & $z_{23}$ & $A$ \\
  \hline
1  & 0    & 0    & 24 & 0    & 0    & 0    & 24 & 19 & 0    & 24 & 0    & 0    & 0    & 0    & 24 \\
2  & 0    & 76 & 0    & 0    & 0    & 0    & 76 & 20 & 0    & 0    & 76 & 0    & 0    & 0    & 76 \\
3  & 24 & 0    & 76 & 0    & 1    & 0    & 100 & 21 & 24 & 76 & 0    & 1    & 0    & 0    & 100 \\
4  & 24 & 0    & 76 & 0    & -1   & 0    & 100 & 22 & 24 & 76 & 0    & -1   & 0    & 0    & 100 \\
5  & 76 & 24 & 0    & -1   & 0    & 0    & 100 & 23 & 076 & 0    & 24 & 0    & -1   & 0    & 100 \\
6  & 76 & 24 & 0    & 1    & 0    & 0    & 100 & 24 & 76 & 0    & 24 & 0    & 1    & 0    & 100 \\
7  & 0    & 0    & 24 & 0    & 0    & 0    & 24 & 25 & 0    & 76 & 24 & 0    & 0    & -1   & 100 \\
8  & 0    & 24 & 76 & 0    & 0    & 1    & 100 & 26 & 0    & 76 & 24 & 0    & 0    & 1    & 100 \\
9  & 0    & 24 & 76 & 0    & 0    & -1   & 100 & 27 & 0    & 0    & 76 &      & 0    & 0    & 76 \\
10 & 24 & 76 & 0    & 1    & 0    & 0    & 100 & 28 & 24 & 0    & 0    & 0    &  0   & 0    & 24 \\
11 & 24 & 76 & 0    & -1   & 0    & 0    & 100 & 29 & 76 & 24 & 0    & -1   & 0    & 0    & 100 \\
12 & 76 & 0    & 0    & 0    & 0    & 0    & 76 & 30 & 76 & 24 & 0    & 1    & 0    & 0    & 100 \\
13 & 25 & 25 & 25 & 1    & 1    & 1    & 75 & 31 & 25 & 25 & 25 & 1    & 1    & 1    & 75 \\
14 & 25 & 25 & 25 & 1    & 1    & -1   & 75 & 32 & 25 & 25 & 25 & 1    & 1    & -1   & 75 \\
15 & 25 & 25 & 25 & 1    & -1   & -1   & 75 & 33 & 25 & 25 & 25 & 1    & -1   & -1   & 75 \\
16 & 25 & 25 & 25 & -1   & 1    & 1    & 75 & 34 & 25 & 25 & 25 & -1   & 1    & 1    & 75 \\
17 & 25 & 25 & 25 & -1   & -1   & 1    & 75 & 35 & 25 & 25 & 25 & -1   & -1    & 1   & 75 \\
18 & 25 & 25 & 25 & -1   & -1   & -1   & 75 & 36 & 25 & 25 & 25 & -1   & -1    & -1  & 75 \\
\hline
\end{tabular}
\end{table}
Once we analyze the above design, we notice that the maximum prediction variance is 0.883, which is not much higher. This indicates that there are some regions in the design space where the model’s prediction may be less reliable. A value of 0.472 for average prediction variance suggests that, on average, model predictions are reasonably precise. A G-efficiency value of 53.5\% indicates that the design is only moderately efficient in terms of prediction throughout the entire design space. The high D-efficiency value 9411.03 suggests that the design is highly efficient in maximizing the determinant of the information matrix, which generally correlates with good overall prediction capability.
\begin{figure}[htbp]
    \centering
    \includegraphics[width=0.6\linewidth]{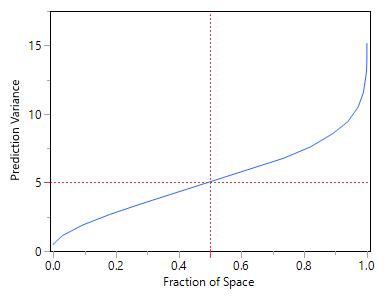}  
    \caption{Fraction of Design Space Plot for OofA component-amount design in orthogonal blocks.}
    \label{fig:FDS-OofA-comp}  
\end{figure}
The FDS plot in Figure \ref{fig:FDS-OofA-comp}  indicates strong prediction capabilities across most of the design space. A low prediction variance over a large fraction of the design space indicates a well-balanced design that provides reliable predictions across different points.
\begin{table}[htbp]
\centering
\caption{Power at 5\% alpha level for effects.}
\label{Tab:9}
\begin{tabular}{cccccccc}
\toprule
Terms & Std. Error & $R_i^2$ & 2 Std. Dev & Terms & Std. Error & $R_i^2$ & 2 Std. Dev \\
\midrule
$a_1$       & 1.09 & 0.9484 & 14.1\% & $a_3^2$      & 1.41 & 0.8853 & 27.0\% \\
$a_2$       & 1.09 & 0.9484 & 14.1\% & $a_1a_2$     & 0.82 & 0.8145 & 21.1\% \\
$a_3$       & 0.97 & 0.9384 & 16.5\% & $a_1a_3$     & 0.96 & 0.8542 & 16.8\% \\
$z_{12}$    & 0.28 & 0.3816 & 91.5\% & $a_2a_3$     & 0.96 & 0.8542 & 16.8\% \\
$z_{13}$    & 0.38 & 0.5667 & 70.5\% & $a_1z_{12}$  & 0.56 & 0.3779 & 39.8\% \\
$z_{23}$    & 0.38 & 0.5667 & 70.5\% & $a_2z_{23}$  & 0.75 & 0.5648 & 24.4\% \\
$a_1^2$     & 1.38 & 0.8664 & 28.1\% & $a_1z_{13}$  & 0.75 & 0.5648 & 24.4\% \\
$a_2^2$     & 1.38 & 0.8664 & 28.1\% & ---          & ---  & ---    & --- \\
\bottomrule
\end{tabular}
\end{table}

The power analysis in Table~\ref{Tab:9} suggests that the components $a_1$, $a_2$, and $a_3$ are less likely to show significant effects at 2 Std. Dev., with power between 14.1\% to 16.5\%. Whereas the high value of $R_i^2$ (0.9384 to 0.9484) suggests that these effects explain a significant portion of the variability in the response. The PWO variables $z_{12}$, $z_{13}$, $z_{23}$ have high power at 2 Std. Dev. (70.5\% to 91.5\%), suggesting they are more likely to be significant in the experiment. The interaction effects $a_1z_{12}$ , $a_2z_{23}$ and $a_1z_{13}$  show moderate significance, with power at 2 Std. Dev. in the range of 24.4\% to 39.8\%, indicating they might also contribute significantly to the response. This reveals that the order-of-addition of components, Sumatriptan, Lactose Monohydrate, and Microcrystalline Cellulose, in Migraine tablet formulation has a significant effect on the crushing strength of the tablet. 
\section{Conclusion}\label{conclusion}
In this study, we defined OofA mixture and mixture-amount experiments in orthogonal blocks. The mixture proportions or the component amounts in orthogonal blocks, and the order of their addition affect the overall response. We modified mixture effect models, including the orthogonal block and order-of-addition effects. We used Scheffe’s quadratic mixture model and K-model for orthogonal block mixture experiments. The orthogonal blocking conditions for these mixture models are already available in the literature. As we incorporated the PWO variables effect in the model containing the orthogonal block effect, we observed that these PWO variables also hold orthogonal blocking conditions defined in section \ref{models_for_OofA_mixture_experiment_in_orthogonal_blocks}.

In the first example of yarn forming in a textile factory, the analysis of the orthogonal block mixture experiment demonstrates that both the proportions of the three key components (Polyethylene, Polystyrene, and Polypropylene) and the order of their addition significantly affect the strength of the yarn produced.  We fitted Scheffé’s quadratic mixture model with orthogonal blocks and order-of-addition effects. The inclusion of order-of-addition factors in the model shows a pronounced impact on the response variable, exhibiting higher significance than the mixture components alone. The interaction terms also reveal moderate influence, indicating the relevance of mixture-order interactions. Overall, the design maintains a balance between prediction accuracy and efficiency, evidenced by the G-efficiency of 58.8\%. 

The second example explores the vinyl production experiment that highlights the importance of both mixture proportions and the order of addition of plasticizers in determining the thickness of the final product. The OofA orthogonal block mixture design given in Table~\ref{Tab:4} was used for fitting the OofA quadratic K-model. The analysis shows that all quadratic mixture effects and order-of-addition variables are statistically significant, suggesting that mixture composition and the sequence of component blending have a notable impact on vinyl thickness. Despite moderate significance in their interactions, the overall model exhibits stable predictive accuracy, with a G-efficiency of 59\%, ensuring reasonable design efficiency. This demonstrates the critical role of both mixture components and processing conditions, such as drying temperature, in optimizing vinyl production outcomes.

Finally, in Example 3, we discussed the formulation and analysis of Migraine relief tablets, taking into account the single process variable granulation method. As the amount of tablet composition and the order of addition of components affect the crushing strength of the tablet, we used an OofA component-amount design in two orthogonal blocks. The analysis of the Migraine relief tablet demonstrates that the order-of-addition of key components (Sumatriptan, Lactose Monohydrate, and Microcrystalline Cellulose) has a significant impact on the crushing strength of the tablet. The design, which incorporates PWO variables and block effects, shows a balanced prediction capability and reasonably precise model predictions. The design's G-efficiency of 53.5\%  suggests moderate prediction efficiency across the design space. The power analysis further emphasizes that the PWO variables ($z_{12}$, $z_{13}$, $z_{23}$) are likely to be significant, whereas the components themselves have lower significance. While moderately significant, the interaction effects still contribute to the response, reinforcing the importance of considering both mixture proportions and the sequence of component addition in optimizing the tablet's physical properties.

The developed OofA mixture designs in orthogonal blocks provide a systematic approach to including order-of-addition effects in mixture and mixture-amount experiments with orthogonal blocks. This enables a more comprehensive design space exploration and better optimization of the response variable. In some research scenarios, F-square orthogonal block designs are preferred over Latin square designs despite requiring more runs.

\section*{Acknowledgements}
Touqeer Ahmad acknowledges support from the R\`egion
Bretagne through project SAD-2021- MaEVa.

\section*{Conflict of interest}
The authors declare no conflict of interest.

\bibliographystyle{plainnat} 
\bibliography{bibliography} 

\begin{thebibliography}{25}
\providecommand{\natexlab}[1]{#1}
\providecommand{\url}[1]{\texttt{#1}}
\expandafter\ifx\csname urlstyle\endcsname\relax
  \providecommand{\doi}[1]{doi: #1}\else
  \providecommand{\doi}{doi: \begingroup \urlstyle{rm}\Url}\fi

\bibitem[Aggarwal et~al.(2004)Aggarwal, Singh, and Gupta]{aggarwal2004orthogonal}
ML~Aggarwal, Poonam Singh, and Nidhi Gupta.
\newblock Orthogonal block designs in two blocks for second degree k-model.
\newblock \emph{Statistics \& Probability Letters}, 66\penalty0 (4):\penalty0 423--434, 2004.

\bibitem[Cornell(2002)]{cornell2002experiments}
John~A. Cornell.
\newblock \emph{Experiments with Mixtures: Designs, Models, and the Analysis of Mixture Data}.
\newblock John Wiley \& Sons, New York, USA, 3rd edition, 2002.

\bibitem[Czitrom(1988)]{czitrom1988mixture}
Veronica Czitrom.
\newblock Mixture experiments with process variables: D-optimal orthogonal experimental designs.
\newblock \emph{Communications in Statistics-Theory and Methods}, 17\penalty0 (1):\penalty0 105--121, 1988.

\bibitem[Ding et~al.(2015)Ding, Matsuo, Xu, Yang, and Zheng]{ding2015optimized}
Xianting Ding, Kyle Matsuo, Lin Xu, Jian Yang, and Longpo Zheng.
\newblock Optimized combinations of bortezomib, camptothecin, and doxorubicin show increased efficacy and reduced toxicity in treating oral cancer.
\newblock \emph{Anti-cancer drugs}, 26\penalty0 (5):\penalty0 547--554, 2015.

\bibitem[Draper and Pukelsheim(1998)]{draper1998mixture}
Norman~R Draper and Friedrich Pukelsheim.
\newblock Mixture models based on homogeneous polynomials.
\newblock \emph{Journal of Statistical Planning and Inference}, 71\penalty0 (1-2):\penalty0 303--311, 1998.

\bibitem[Fisher(1971)]{fisher1971design}
R.~A. Fisher.
\newblock \emph{The Design of Experiments}.
\newblock Macmillan, London, UK, 9th edition, 1971.

\bibitem[Jiang and Ng(2014)]{jiang2014sequential}
Xiong-Jie Jiang and Dennis~KP Ng.
\newblock Sequential logic operations with a molecular keypad lock with four inputs and dual fluorescence outputs.
\newblock \emph{Angewandte Chemie International Edition}, 53\penalty0 (39):\penalty0 10481--10484, 2014.

\bibitem[John(1984)]{john1984experiments}
Peter~WM John.
\newblock Experiments with mixtures involving process variables.
\newblock \emph{Statistics Department Technical Report}, \penalty0 (8), 1984.

\bibitem[Karim et~al.(2000)Karim, McCormick, and Kappagoda]{karim2000effects}
Malina Karim, Kellie McCormick, and C~Tissa Kappagoda.
\newblock Effects of cocoa extracts on endothelium-dependent relaxation.
\newblock \emph{The Journal of Nutrition}, 130\penalty0 (8):\penalty0 2105S--2108S, 2000.

\bibitem[Lin and Peng(2019)]{lin2019order}
Dennis~KJ Lin and Jiayu Peng.
\newblock Order-of-addition experiments: A review and some new thoughts.
\newblock \emph{Quality Engineering}, 31\penalty0 (1):\penalty0 49--59, 2019.

\bibitem[Munija and Srikanth(2018)]{munija2018formulation}
P~Munija and G~Srikanth.
\newblock Formulation and evaluation of sumatriptan immediate release tablets.
\newblock \emph{Journal of Drug Delivery and Therapeutics}, 8\penalty0 (5):\penalty0 241--247, 2018.

\bibitem[Murty(1966)]{murty1966}
J.~S. Murty.
\newblock \emph{Problems of Construction and Analysis of Designs of Experiments}.
\newblock PhD thesis, Delhi University, 1966.
\newblock Unpublished PhD thesis.

\bibitem[Nigam(1970)]{nigam1970block}
AK~Nigam.
\newblock Block designs for mixture experiments.
\newblock \emph{The Annals of Mathematical Statistics}, 41\penalty0 (6):\penalty0 1861--1869, 1970.

\bibitem[Nigam(1976)]{nigam1976correction}
AK~Nigam.
\newblock Correction to" block designs for mixture experiments".
\newblock \emph{The Annals of Statistics}, 4\penalty0 (6):\penalty0 1294--1295, 1976.

\bibitem[Peng et~al.(2019)Peng, Mukerjee, and Lin]{peng2019design}
Jiayu Peng, Rahul Mukerjee, and Dennis~KJ Lin.
\newblock Design of order-of-addition experiments.
\newblock \emph{Biometrika}, 106\penalty0 (3):\penalty0 683--694, 2019.

\bibitem[Piepel(1988)]{piepel1988note}
Gregory~F Piepel.
\newblock A note on models for mixture-amount experiments when the total amount takes a zero value.
\newblock \emph{Technometrics}, 30\penalty0 (4):\penalty0 449--450, 1988.

\bibitem[Piepel and Cornell(1985)]{piepel1985models}
Gregory~F Piepel and John~A Cornell.
\newblock Models for mixture experiments when the response depends on the total amount.
\newblock \emph{Technometrics}, 27\penalty0 (3):\penalty0 219--227, 1985.

\bibitem[Prescott and Draper(2004)]{prescott2004mixture}
Philip Prescott and Norman~R Draper.
\newblock Mixture component-amount designs via projections, including orthogonally blocked designs.
\newblock \emph{Journal of Quality Technology}, 36\penalty0 (4):\penalty0 413--431, 2004.

\bibitem[Rios and Lin(2022)]{rios2022order}
Nicholas Rios and Dennis~KJ Lin.
\newblock Order-of-addition mixture experiments.
\newblock \emph{Journal of Quality Technology}, 54\penalty0 (5):\penalty0 517--526, 2022.

\bibitem[Scheff{\'e}(1958)]{scheffe1958experiments}
Henry Scheff{\'e}.
\newblock Experiments with mixtures.
\newblock \emph{Journal of the Royal Statistical Society: Series B (Methodological)}, 20\penalty0 (2):\penalty0 344--360, 1958.

\bibitem[Scheffe(1963)]{scheffe1963simplex}
Henry Scheffe.
\newblock The simplex-centroid design for experiments with mixtures.
\newblock \emph{Journal of the Royal Statistical Society: Series B (Methodological)}, 25\penalty0 (2):\penalty0 235--251, 1963.

\bibitem[Van~Nostrand(1995)]{van1995design}
RC~Van~Nostrand.
\newblock Design of experiments where the order of addition is important.
\newblock In \emph{ASA proceedings of the Section on Physical and Engineering Sciences}, volume 155, page 160. American Statistical Association Alexandria, VA, USA, 1995.

\bibitem[Voelkel(2019)]{voelkel2019design}
Joseph~G Voelkel.
\newblock The design of order-of-addition experiments.
\newblock \emph{Journal of Quality Technology}, 51\penalty0 (3):\penalty0 230--241, 2019.

\bibitem[Winker et~al.(2020)Winker, Chen, and Lin]{winker2020construction}
Peter Winker, Jianbin Chen, and Dennis~KJ Lin.
\newblock The construction of optimal design for order-of-addition experiment via threshold accepting.
\newblock In \emph{Contemporary Experimental Design, Multivariate Analysis and Data Mining: Festschrift in Honour of Professor Kai-Tai Fang}, pages 93--109. Springer, Cham, 2020.

\bibitem[Yuna~Zhao and Liu(2021)]{Zhao2021}
Dennis K. J.~Lin Yuna~Zhao and Min-Qian Liu.
\newblock Designs for order-of-addition experiments.
\newblock \emph{Journal of Applied Statistics}, 48\penalty0 (8):\penalty0 1475--1495, 2021.

\end{thebibliography}

\end{document}